\def\be{\begin{equation}}
\def\ee{\end{equation}}
\def\te{\end{equation}}
\def\bea{\begin{eqnarray}}
\def\ba{\begin{eqnarray}}

\def\ta{\end{eqnarray}}
\def\tea{\end{eqnarray}}
\def\eea{\end{eqnarray}}

\def\ben{\begin{enumerate}}
\def\een{\end{enumerate}}

\documentclass[aps,prd,amssymb,amsmath,nofootinbib]{revtex4}

\usepackage[export]{splitbib}

\begin{category}[A]{Stochastic gravity} 
\SBentries{stogra,stograRev,CH87,CamVer94,HuPhysica,CH94,%
HM3,fdrscg,CamVer96,ELE,MV1,MV2} 
\end{category} 
\begin{category}[B]{Noise kernel} 
\SBentries{PH1,PH2,EHR} 
\end{category}
\begin{category}[C]{Black hole in thermal radiance} 
\SBentries{GibPer,BHinBox,HawPage,Page82,WY,BroYor,BraYor,%
HKM,BekMei,Schiffer,PanWald77,Zur80} 
\end{category}
\begin{category}[D]{Black hole back-reaction from radiance} 
\SBentries{CanSci,Mottola,York,York85,AHS,HLA,2DilatonGR,%
CamHu,SRH}
\end{category}
\begin{category}[E]{Evaporating black hole} 
\SBentries{Bar81,Israel,Hiscock,Massar} 
\end{category}
\begin{category}[F]{Black hole horizon fluctuations} 
\SBentries{Bek84,Casher,Sorkin,Marolf,HRS,ForSva,WuFor99,%
MasPar,BFP,Par02,HRevapBHfluc,HRevapBHlong} 
\end{category}

\begin{document}

\title{Black hole fluctuations and dynamics from back-reaction of
Hawking radiation: Current work and further studies based on
stochastic gravity}
\author{B.~L. Hu$^1$\footnote{Email address: hub@physics.umd.edu}
and Albert Roura$^{1,2}$\footnote{Email address: roura@lanl.gov}}
\affiliation{$^1$Department of Physics, University of Maryland,
College Park, Maryland 20742-4111 \\
$^2$Theoretical Division, T-8, Los Alamos National Laboratory, M.S.
B285, Los Alamos, NM 87545 }

\begin{abstract}
We give a progress report of our research on spacetime fluctuations
induced by quantum fields in an evaporating black hole and a black
hole in quasi-equilibrium with its Hawking radiation. We note the
main issues involved in these two classes of problems and outline
the key steps for a systematic quantitative investigation. This
report contains unpublished new ideas for further studies.
\end{abstract}

\maketitle

\setcounter{page}{1}
%\newpage
\section{Black hole back-reaction and fluctuations: Our current work}

Back-reaction here refers to the influence of particle creation as
in the Hawking radiation and other quantum effects (such as trace
anomaly) on the structure and dynamics of the background spacetime.
It is believed to grow in importance when the energy reaches the
Planck scale, as in the very early universe and at the final stages
of black hole evolution. In semiclassical black hole physics this is
perhaps one of the more difficult but important unsolved problems.
It is important because its solution is necessary for us to
understand better the black hole end-state and information loss
puzzles. It also provides a check on the range of validity for
Hawking's derivation of black hole radiance in the framework of
semiclassical gravity. As we found out from the \textsl{stochastic
gravity} perspective \cite{stogra,stograRev}, the back-reaction
problem is tied to dissipation and fluctuation phenomena. Thus it
provides a natural generalization to \textsl{non-equilibrium black
hole thermodynamics} and can reveal deeper connections between
gravity, quantum field theory and statistical mechanics.

Difficulties in the black hole back-reaction problems start with
finding a renormalized energy momentum tensor\footnote{For a list of
papers dedicated to this task, see footnote 1 of \cite{SRH}. Some
notable work under the framework of semiclassical gravity (in
addition to those few discussed in more detail below) includes
Anderson \emph{et al.} \cite{AHS} for a quantum scalar field in a
spherically symmetric spacetime, the back-reaction in the interior
of a black hole \cite{HLA} (easier technically since it is a
cosmological Kantowski-Sachs spacetime) and for two-dimensional
dilaton gravity theory \cite{2DilatonGR}.}. The stochastic gravity
program \cite{stograRev}  introduces fresh insight and new
methodology into the back-reaction problem by a) stressing the
importance of an in-in (closed-time-path, or CTP) formulation
\cite{CH87,CamVer94} which gives real and causal equations of
motion, b) imparting a statistical mechanical meaning to the
back-reaction effects via quantum open system concepts and
techniques \cite{HuPhysica,CH94}, c) noting the necessity of
including fluctuations in conjunction with dissipation in the system
dynamics and showing how noise arises with the help of the influence
functional formalism \cite{HM3,fdrscg,CamVer96}.  The quantities of
importance are the dissipation kernel, which enters into the
Einstein-Langevin (E-L) equation \cite{ELE}, and the noise kernel,
which characterizes the correlations of its stochastic source.

The back-reaction problems of interest to us fall into two classes:
1) A black hole in a box in quasi-equilibrium with its Hawking
radiation. 2) An evaporating black hole emitting Hawking radiation
under fully non-equilibrium conditions.  These problems have been
treated before in varying degrees of completeness. The former is
easier to understand but has some remaining technical challenges. We
approach this problem at two levels: i) At the \textit{quantum field
theory level} is the derivation of the influence action, which
demands a computation of the noise kernel for quantum fields near
the Schwarzschild horizon.
% This requires a higher level of technical skills beyond that for the stress energy tensor.
ii) At the \textit{statistical thermodynamic level} we approach this
problem by viewing the back-reaction as an embodiment of the
fluctuation-dissipation relation (FDR). We have demonstrated the
usefulness of this way of thinking in cosmological back-reaction
problems \cite{fdrscg}.
%and will apply the insight gained to the present black hole back-reaction problems.
The problems for evaporating black holes have only been treated
sparsely and qualitatively, even with contradictory claims. Our
current effort has focused on clarifying some existing conceptual
confusion and building a unifying framework capable of producing more
quantitative results.

\section{Quasi-equilibrium conditions: Black hole in a box}

A black hole can be in quasi-equilibrium with its Hawking radiation
if placed inside a box of the right size
\cite{BHinBox,WY}\footnote{Note that one needs to introduce the
appropriate redshift factors that account for the finite size of the
box. Moreover, for a sufficiently small box, as required in general
to have (meta)stable equilibrium, curvature corrections, which were
not included in Ref.~\cite{CamHu} may not be negligible.}, or in an
anti-de Sitter universe \cite{HawPage}. We divide our consideration
into the far-field case and the near-horizon case. The far field
case has been studied before \cite{CamHu}. In \cite{SRH} we consider
the near-horizon case. Using the model of a black hole described by
a radially-perturbed quasi-static metric and Hawking radiation by a
conformally coupled massless quantum scalar field, we showed that
the closed-time-path (CTP) effective action yields a non-local
dissipation term as well as a stochastic noise term in the equation
of motion.
%When the noise average of this Einstein-Langevin equation is taken, we recover York's
%\cite{York} semiclassical equations for radially-perturbed quasi-static black holes.
We have presented the overall structure of the theory and the
strategy of our approach in \cite{SRH}, but due to the lack of an
analytic form of the Green function for a scalar field in the strong
field region of the Schwarzschild metric, numerical calculations may
be the only way to go. Being based on a quasilocal expansion, Page's
approximation \cite{Page82}, though reasonably accurate for the
stress tensor, is insufficient for the noise kernel since one needs
to consider arbitrarily separated points in that case. In \cite{SRH}
we also presented an alternative derivation of the CTP effective
action in terms of the Bogoliubov coefficients, thus connecting with
the interpretation of the noise term as measuring the difference in
particle production in alternative histories. [This will be useful
for a grand-canonical ensemble description of black hole
fluctuations. (See below.)]

\subsection{Connecting different approaches}

On the black-hole-in-a-box back-reaction problem we take a two-prong
approach: via quantum field theory and statistical thermodynamics.
We are currently performing a calculation of the noise kernel near
the Schwarzschild horizon, making use of the results of \cite{PH2}
but with points kept separate \cite{EHR}. We will use the result of
this calculation to show the existence of a fluctuation-dissipation
relation (described below). In addition we want to integrate the
master equation approach of Zurek \cite{Zur80} and the transition
probability approach of Bekenstein, Meisel \cite{BekMei} and
Schiffer \cite{Schiffer} (see also \cite{PanWald77}) under the
stochastic gravity framework. Within the thermodynamic descriptions
we want to compare the results from the canonical with the
microcanonical \cite{BroYor} and the grand canonical ensembles
\cite{BraYor}\footnote{Note, however, that for massless particles
with vanishing chemical potential the canonical and grand canonical
ensembles coincide}. For the last task we will invoke results
obtained earlier \cite{HKM} for the number fluctuations in particle
creation and the relation we obtained recently in \cite{SRH}
expressing the CTP effective action in terms of the Bogoliubov
coefficients for the black hole particle creation, to derive the
susceptibility and isothermal compressibility functions of the black
hole. This would move us a step closer to establishing a linear
response theory (LRT) of non-equilibrium black hole thermodynamics.

\subsection{Back-reaction manifested through a
fluctuation-dissipation relation}

Historically Candelas and Sciama \cite{CanSci} first suggested a
fluctuation-dissipation relation  for the depiction of dynamic black
hole evolution. They proposed a classical formula relating the
dissipation in area linearly to the squared absolute value of the
shear amplitude. The quantized gravitational perturbations (they
choose the quantum state to be the Unruh vacuum) approximate a flux
of radiation emanating from the hole at large radii. Thus they claim
that the dissipation in area due to the Hawking flux of
gravitational radiation is related to the quantum fluctuations of
the metric perturbations. However, as was pointed out in \cite{HRS},
it is not clear that their relation corresponds to a FDR in the
correct statistical mechanical sense and does not include the effect
of matter fields. The FDR for the contribution of the matter fields
should involve the fluctuations of the stress tensor (the
``generalized force'' acting on the spacetime), which are
characterized by the noise kernel. With an explicit calculation of
the noise kernel in a similar context one could obtain the correct
FDR.
%But as was pointed out in \cite{HRS}, their relation does
%not represent back-reaction accurately and is not a FDR in the
%correct statistical mechanical sense. Specifically, they derive an
%expression for the dissipation in the black hole horizon area due to
%the Hawking flux of gravitational radiation, and relate it to the
%quantum fluctuations of gravitons. The correct entry in a FDR should
%be the fluctuations of the area, which is related to the two point
%function of the stress tensor; the true ``generalized force'' acting
%on the spacetime is indeed the noise kernel. With an explicit
%calculation of the noise kernel in a similar context one could
%obtain the correct FDR.

For the quasi-static case Mottola \cite{Mottola} introduced a FDR
based on linear response theory  for the description of a black hole
in quasi-equilibrium with its Hawking radiation. He showed that in
some generalized Hartle-Hawking state a FDR
exists between the expectation values of the
commutator and anti-commutator of the energy-momentum tensor.  This
is the standard form of FDR. However, in his case the dynamical
equation for the linear metric perturbations describes just their
mean evolution, which corresponds to taking the stochastic average
of the linearized E-L equation. The E-L equation from stochastic
gravity does more in providing the dynamics of the fluctuations.
%However, linear response theory in the
%way usually presented assumes a fixed finite temperature bath and
%examines how a system reacts to an external perturbance in the
%linear regime. This formulation does not give a complete description
%of back-reaction \cite{HRS} because it is based on a specific
%background spacetime (static in this case) and state (thermal) of
%the matter field(s). In the more complete and correct treatment of
%the back-reaction problem,  the spacetime and the state of matter
%should be determined in a self-consistent manner by their dynamics
%and mutual influence. For example, the linear response dissipation
%kernel is given by a two-point commutator function of the underlying
%quantum field, whereas in truth it should depend on a two-point
%function of the stress-tensor, which is the noise kernel which is
%a four-point function of the field. %is in general state-dependent.

In a recent paper \cite{SRH} we laid out the road map for treating
the quasi-static case in the stochastic gravity framework and point
out that a non-local dissipation and a fluctuation term will arise
which should match with the analytic results in the far field limit
derived earlier \cite{CamHu}. These terms are absent in York's work
\cite{York} because the approximate form he uses for the
semiclassical Einstein equation  corresponds to the variation of
terms in the CTP effective action which are linear in the metric
perturbations around the Schwarzschild background geometry, whereas
the non-local dissipation and noise kernels appear at quadratic
order. The noise kernel, which is connected to the fluctuations of
the stress tensor, would give no contribution to his equation for
the mean evolution, even if higher order corrections were included.
%since his result describes the correction to the
%Schwarzschild geometry with only radial perturbations due to the
%presence of a quantum test field in that geometry, with the
%resultant geometry remaining {static}.  The non-local terms arise
%when considering the equation for linear perturbations around such a
%background spacetime as a self-consistent solution to the
%semiclassical Einstein equation.
%This issue is currently being explored along these directions for these
%two classes of problems.

%%%%%%%%%%%%%%%%%%%%%%%%%%%%%%%%%%%%%%%%%%%%%%%%%%%%%%%%%%%%%%%%%%%%%

\section{Evaporating black hole: Non-equilibrium conditions}

This time-dependent problem has a very different physics from the
quasi-equilibrium case, and is in general more difficult to treat.
Starting in the early 80's, it has been approached by Bardeen
\cite{Bar81}, Israel \cite{Israel} and Massar \cite{Massar}, who
considered the mean evolution. The role of fluctuations was initially
studied by Bekenstein \cite{Bek84} and has received further attention
in recent years by Ford \cite{ForSva,WuFor99}, Frolov \cite{BFP},
Sorkin \cite{Sorkin}, Marolf \cite{Marolf}, and their collaborators
(see also Ref.~\cite{Casher}), largely based on qualitative
arguments. On some issues, such as the size of black hole horizon
fluctuations, there are contradictory claims.

To make progress one needs to introduce a theoretical framework
where all prior claims can be scrutinized and compared. Because of
its non-equilibrium nature we expect the stochastic gravity program
\cite{stogra} to provide some useful quantitative results. In
\cite{HRS} we wrote down the Einstein-Langevin equation for the
fluctuations in the mass of an evaporating black hole and found that
the fluctuations compared to the mean is small unless the mean
solution is unstable with respect to small perturbations. Our 1998
paper emphasized the first part of this statement, which is in
apparent contradiction to what Bekenstein claimed in his 1984 paper
\cite{Bek84}.  Recently we revisited this question with a closer
analysis \cite{HRevapBHfluc}.  Since for an evaporating black hole
there exist unstable perturbations around the solution of the
semiclassical Einstein equation for the mean evolution, the second
part of the statement applies.

A key assumption in these studies is that the fluctuations of the
incoming energy flux near the horizon are directly related to those
of the outgoing one (a condition that does hold for the mean flux).
If this condition were true, one can indeed show that the
fluctuations of a black hole horizon become important (growing
slowly, but accumulating over long times), as Bekenstein claimed.
However, we have serious reservations on its validity for energy
flux fluctuations (see below). Using the E-L equation we also point
out how the different claims of Bekenstein and Ford-Wu can be
reconciled by recognizing the different physical assumptions they
used in their arguments. The stochastic gravity theory which our
present work is based on should provide a platform for further
investigations into this important issue.

%%%%%%%%%%%%%%%%%%%%%%%%%%%%%%%%%%%%%%%%%%%%%%%%%%%%%%%%%%%%%%
%\section{Current and Future work}
%%%%%%%%%%%%%%%%%%%%%%%%%%%%%%%%%%%%%%%%%%%%%%%%%%%%%%%%%%%%%%

%We will continue our current program  to treat two classes of black
%hole back-reaction and fluctuation problems: On the evaporating black
%hole problem we will clarify the confusion from different approaches
%in the literature and show how they can be unified under the
%stochastic gravity program. We then derive the dynamics of the
%fluctuations using the Einstein-Langevin equation which to us seems
%to be the best method to generate quantitative results for this problem.

\subsection{Bardeen's evaporating black hole and Bekenstein's
fluctuation theory}

In 1981 Bardeen \cite{Bar81} considered the back-reaction of Hawking
radiation on an evaporating black hole by invoking a Vaidya-type
metric. (This model was later used by Hiscock \cite{Hiscock} for
similar inquiries.) Bardeen's calculation with this model affirmed the
validity of the semiclassical picture assumed in Hawking's original
derivation of thermal radiance. His results including back-reaction
indicate that the black hole follows an evolution which is largely
determined by the semiclassical Einstein equations down to where the
black hole mass drops to near the Planck mass ($\sim 10^{-5}$ g), the
point where most practitioners of semiclassical gravity would agree
that the theory will break down. Bardeen's result was developed
further by Massar \cite{Massar}.

For black hole mass fluctuations, in 1984 Bekenstein \cite{Bek84}
considered the mass fluctuations of an {\it isolated} black hole due
to the energy fluctuations in the radiation emitted by the hole, and
asks when such fluctuations become large. According to his
calculation, depending on the initial mass, mass fluctuations can
grow large well before the mass of the black hole reaches the Planck
scale.
%Once a critical mass is reached, the
%evolution of the black hole differs considerably from that predicted
%by Bardeen and Massar. Of the few work in this setting, the result
%of Wu and Ford \cite{WuFor99} supports Bardeen's scenario.
On the other hand, of the few studies in this setting, the result of
Wu and Ford \cite{WuFor99} supports a scenario in which fluctuations
do not become important before reaching the Planckian regime.  In
view of such contradictory claims in the literature, it is highly
desirable to have a more solid and complete theoretical framework
where all prior claims can be scrutinized and compared. We expect
the stochastic gravity program to be useful for this purpose. At the
most rudimentary level, Bekenstein's approach shares similar
conceptual emphasis as does the stochastic gravity program in that
both attribute importance to the fluctuations of stress tensor and
the black hole mass, characterized by their correlation functions.

\subsection{Non-equilibrium conditions}

Investigations in this case may assist in answering two important
questions: {\it 1) Are the fluctuations near the horizon large or
small? 2) How reliable are earlier results from test quantum
fields in fixed curved spacetime?}

As far as the first question is concerned, one should distinguish
between fluctuations with short and long characteristic time scales.
For time scales comparable to the evaporation time, one expects
spherically-symmetric modes ($l=0$) to be dominant. Moreover, if one
assumes a direct correlation between the energy flux fluctuations on
the horizon and those far from it, as done in earlier work, an
explicit result supporting Bekenstein's conclusion can be obtained
and the origin of the discrepancy with Ford and Wu's result can be
understood. However, a more careful analysis reveals that such an
assumption is not correct (see below) and it is necessary to compute
the noise kernel near the horizon to get an accurate answer. One
can present arguments to the effect that  this kind of fluctuations
may not modify in a drastic way the result obtained by Hawking for
test fields evolving on a fixed black hole spacetime, and later
extended by Bardeen and Massar to include their back-reaction effect
on the mean evolution of the spacetime geometry. On the other hand,
in principle,  fluctuations with much shorter correlation times,
which also involve higher multipoles ($l \neq 0$), could alter
substantially Hawking's result.

Detailed calculations within the framework of stochastic gravity can
address those issues in a natural way, at least for fluctuations
with typical scales much larger than the Planck length.
Nevertheless, one needs to pay attention to the subtleties.
Preliminary results, briefly described below, signal a possible
breakdown of the geometric optics approximation for the propagation
of test fields when fluctuations are included. This would require
finding alternative ways of probing satisfactorily those metric
fluctuations and extracting physically meaningful information.

\subsubsection{Spherically symmetric sector ($l = 0$)}

Consideration of this problem with restriction to spherically
symmetric modes was done in \cite{Zur80,Bek84,WuFor99,BFP,Par02} as
well as two-dimensional dilaton-gravity models \cite{2DilatonGR}.
All those studies restricted from the outset the contribution to the
classical action to $s$-wave modes for both the metric and the
matter fields, which only allows $l=0$ modes for each matter field
to contribute to the noise kernel.  Our approach goes well beyond
that approximation since modes with all possible values of $l$ for
the matter fields can contribute to the $l = 0$ sector of the noise
kernel.

We focus on the physics in the \textit{adiabatic regime}, which is
the time when the black hole mass $M$ remains much larger than the
Planck mass $m_\mathrm{p}$. This allows one to use $\langle
\hat{T}_{ab} \rangle_\mathrm{ren}$ in the Schwarzschild spacetime
for each instant of time but with a mass slowly evolving in time.
Technically this saves one the trouble of solving the
integro-differential (semiclassical Einstein) equation to obtain the
mean evolution.

In the adiabatic regime, one can show from energy-momentum
conservation that the outgoing (for $r \gg 2M$) and ingoing (for $r
\approx 2M$) \textit{mean fluxes} are related. If a similar argument
is employed to relate the outgoing and ingoing energy fluxes for the
\textit{stochastic source} characterizing the stress tensor
fluctuations, one can proceed in the same way as was done for the
mean fluxes to provide a justification for Bekenstein's approach.
%From the Einstein-Langevin equation, after imposing spherical symmetry,
%one has \be ELeq \ee  This is the starting point of Bekenstein's argument.

This energy conservation argument has been assumed to be valid for
energy flux fluctuations by Bekenstein \cite{Bek84} as well as Wu and
Ford \cite{WuFor99} for an evaporating black hole.
%In the region far from the horizon $\rightarrow$ only need to smear
%the noise kernel in time.
With this assumption one can also clarify the apparent discrepancy
with Wu and Ford as follows. The growth in time of the fluctuations
can be understood in terms of the following ``instability'' exhibited
by the mean evolution: if the initial mass of a macroscopic black hole
with $M \gg m_\mathrm{p}$ is slightly perturbed by a small amount of
the order of the Planck mass, the difference between the masses of the
perturbed and unperturbed black holes becomes of the same order as the
mass of the unperturbed black hole when the latter becomes of order
$(m_\mathrm{p}^2 M)^{1/3}$, \emph{i.e.}, still much larger than the
Planck mass.  This growth of the fluctuations, first found by
Bekenstein, is a consequence of the secular effect of the renormalized
stress energy tensor of the perturbations,
% $\langle \hat{T}_{ab}^{(1)} \rangle_\mathrm{ren}$,
whose effect %, despite being perturbatively smaller than the stochastic source,
builds up in time and gives a contribution to the mass evolution of
the same order as the mean evolution for times of the order of
the black hole evaporation time even when the mass of the black
hole is still much larger than the Planck mass. This term was not
taken into account by Wu and Ford, which explains why they found
much smaller fluctuations than Bekenstein for times of the order
of the evaporation time.\footnote{An interesting related
question is whether decoherence effects render those fluctuations
effectively classical, so that they can be regarded as
fluctuations within an incoherent statistical ensemble rather
than coherent quantum fluctuations.}

Moreover, due to the nonlinear dependence of the flux of radiated
energy on the mass of the black hole, terms of higher order in the
perturbations become relevant when the fluctuations become of the
same order as the mean value of the mass. As pointed out by
Bekenstein, this implies a deviation from the usual semiclassical
Einstein equation for the evolution of the ensemble/stochastic
average of the mass.\footnote{This effect can be interpreted as
follows: due to the growth of the mass fluctuations, the higher
order radiative corrections to the semiclassical equation, which
involve Feynman diagrams with internal lines corresponding to
correlation functions of the metric perturbations (mass
perturbations in this case), can no longer be neglected.}

%{\it Assumption Valid?} If the argument relating the outgoing flux
%and the flux crossing the horizon were true, one could use the
%result for the noise kernel far from the horizon obtained by Wu and
%Ford for an evaporating black hole as described above. For a black
%hole in equilibrium one could use the result for the noise kernel of
%a scalar field in thermal equilibrium in flat space obtained by
%Campos and Hu \footnote{The appropriate redshift factors should be
%included to account for the finite size of the cavity that is needed
%to have (meta)stable equilibrium}.\\

\subsubsection{No correlation between the fluctuations of the energy
flux crossing the horizon and far from it}

%As an explicit example of the application of stochastic gravity, we
%study the spherically-symmetric sector of metric perturbations
%around an evaporating black hole background geometry. For
%macroscopic black holes we find that those fluctuations grow and
%eventually become important when considering sufficiently long
%periods of time (of the order of the evaporation time), but well
%before the Planckian regime is reached. In addition,

At the time of this meeting (Nov. 2005) we had serious doubts that
the same argument that connects the outgoing flux and the flux
crossing the horizon, valid for the expectation value of the stress
tensor, could also hold for the stochastic source accounting for the
stress tensor fluctuations. The reason we gave was the following:
while the time derivative of the expectation value, being of higher
order in $(m_\mathrm{p} / M)$, is negligible in the adiabatic
regime, that is not always the case for the stochastic source.
Therefore, when integrating the conservation equation and computing
the correlation function for the flux crossing the horizon, one gets
additional terms besides the correlation function for the outgoing
flux. Soon after  we came up with a proof that this relation does
not hold. This can be found in \cite{HRevapBHlong}, where the
assumption of a simple correlation between the fluctuations of the
energy flux crossing the horizon and far from it, which was made in
earlier work on spherically-symmetric induced fluctuations, was
carefully analyzed and found to be invalid (see also
Ref.~\cite{Par02} for a related result in an effectively
two-dimensional model).
%
%Our analysis suggests the existence of an
%infinite amplitude for the fluctuations of the horizon as a
%three-dimensional hypersurface. We emphasized the need for
%understanding and designing operational ways of probing quantum
%metric fluctuations near the horizon and extracting physically
%meaningful information.
%
This recent finding would invalidate the working assumption of prior
results on black hole event horizon fluctuations based on
semiclassical gravity, and points to the necessity of doing the
calculation of the noise kernel near the horizon in all seriousness,
an effort barely gotten started a few years ago \cite{PH1,PH2}.

\subsubsection{Non-spherically symmetric sector ($l \neq 0$)}

Sorkin \cite{Sorkin}, Casher \emph{et al.} \cite{Casher} and Marolf
\cite{Marolf} have provided qualitative arguments for the existence
of large quantum fluctuations of the event horizon involving time
scales much shorter than the evaporation time\footnote{In all cases
a Schwarzschild spacetime rather than that of an evaporating black
hole was considered. However, this is a good approximation if one is
interested in analyzing fluctuations with correlation times much
shorter than the evaporation time. Moreover, one expects for similar
reasons that if those large fluctuations did actually exist, they
would also be present in the equilibrium case.}  which would give
rise to an effective width of order
$(R_\mathrm{S}l_\mathrm{p}^2)^{1/3}$, much larger than the Planck
length (in all cases induced rather than intrinsic fluctuations are
implicitly considered). However, Ford and Svaiter \cite{ForSva}
pointed out that Casher \emph{et al.}'s result was probably an
artifact from invoking a wrong vacuum to evaluate the
fluctuations.%
%due to the fact that they considered the fluctuations of the quantum matter field
%with respect to the Boulware vacuum. Whereas the natural state (with a finite
%renormalized expectation value of the stress tensor on the horizon) to consider
%for an eternal black hole is the Hartle-Hawking vacuum (or the Unruh vacuum for an evaporating
%black hole), the Boulware vacuum exhibits large fluctuations with
%respect to the Hartle-Hawking vacuum. It seems, therefore, likely
%that both contributions could interfere destructively and
%essentially cancel out when computing the gravitational effect of
%the stress tensor fluctuations of the Hartle-Hawking vacuum using
%the Boulware vacuum as a starting point.
%Additional support to this argument is provided by
%considering a Rindler horizon in Minkowski spacetime.
\footnote{One expects that a small region near the event horizon
for a very large black hole should be very similar to a Rindler
horizon. The arguments of Casher \emph{et al.} would lead to
large fluctuations (actually infinite in the limit of infinite
radius), but one knows that the fluctuations in Minkowski
spacetime are small.}  Sorkin's result is based on Newtonian
gravity, but
%it is not obvious how one could formulate Ford and Svaiter's objection in that case.
%However, since
Marolf's work is intended to be a generalization of Sorkin's to the
general relativistic case. We intend to clarify these apparently
contradictory claims and treat it with the E-L equation and the
noise kernel calculations for this case.
%the same kind of criticism about Casher \emph{et al.}'s treatment could apply to Marolf's argument.

Additional insight into this problem can be gained by studying induced
metric fluctuations in de Sitter spacetime. A static observer in de
Sitter spacetime perceives the quantum fluctuations of the
Bunch-Davies vacuum as a thermal equilibrium distribution in the same
way a static observer outside a black hole event horizon would
perceive the quantum fluctuations of the Hartle-Hawking vacuum.  The
high degree of symmetry of de Sitter spacetime makes it easier to
obtain exact analytical results. We can check the validity of claims
of large black hole horizon fluctuations by studying the corresponding
problems in the de Sitter spacetime.

\section{Probing metric fluctuations near the horizon}

Our recent finding that there exists no simple connection between
the outgoing flux and the flux crossing the horizon implies that one
needs to compute the noise kernel near the horizon. Ideally one
would compute the noise kernel everywhere in a Schwarzschild
background, but the difficulties mentioned above make it very hard
to obtain an analytical result. On the other hand, having a good
approximation for the noise kernel near the horizon might be enough
to get the main features because the emission of Hawking radiation
is mostly dominated by what is happening near the
horizon.\footnote{The effect from the fluctuations of the potential
barrier for the equation of motion of the radial component could
also be important, but it is not taken into account here.}

\subsection{Computing the noise kernel near the horizon}

The key ingredient to compute the noise kernel in a given background
spacetime and for a given (vacuum) state of the quantum matter
fields is the Wightman function $G^+(x,y)$. Page developed an
approximation to obtain two-point Green functions for spacetimes
which are a vacuum solution of the Einstein equation and are
conformally related to an ultrastatic spacetime. Thus, it can be
applied in particular to the Schwarzschild spacetime. Page's
approximation involves (among other things) an expansion in terms of
the geodetic interval $\sigma$ between the two points in the Green
function starting at order $1/\sigma$ and valid up to order
$\sigma^2$. Page used it to obtain the renormalized expectation
value of the stress tensor operator. The expansion up to order
$\sigma^2$ was enough for his purpose because, after applying the
appropriate differential operators and subtracting the divergent
terms in the renormalization process, the contribution from terms of
order $\sigma^2$ or higher in the expansion of the Green function
vanishes when the coincidence limit is finally taken.

On the other hand, when computing the noise kernel (which involves a
product of two Wightman functions) using Page's approximation for
the Wightman function one obtains an expansion in powers of $\sigma$
starting at order $1/\sigma^4$, which coincides with the flat space
result, and valid through order $1/\sigma$. Furthermore, since the
only additional scale that appears in the problem is the
Schwarzschild radius of the black hole ($= 2M$, the mass in
geometrical/Planckian units), one can conclude by dimensional
analysis that the dimensionless expansion parameter is proportional
to $\sigma/M$. In contrast to the expectation value of the stress
tensor, one does not need to take the coincidence limit $\sigma
\rightarrow 0$ when computing the noise kernel. In fact, when
projecting onto the subset of spherically symmetric multipoles, one
needs to integrate over the whole solid angle for the two points
appearing in the noise kernel. Unfortunately, that involves
considering pairs of points with $\sigma \sim M$, for which Page's
expansion in terms of $\sigma/M$ would break down. However, since
the first few terms contain inverse powers of $\sigma/M$, the
integral of the noise kernel over the whole solid angle is dominated
by the contribution from small angle separations, \emph{i.e.}, pairs
of points with $\sigma/M \ll 1$. Therefore, one expects that using
Page's approximation for the Wightman function when computing the
integral of the noise kernel over the whole solid angle would
provide a fairly good approximation to the actual result.
\footnote{Note, however, that one needs to choose the appropriate
analytical continuation to go from the Euclidean Green function
obtained by Page to the Lorentzian Wightman function. Page's scheme
provides the Green function for Hartle-Hawking vacuum, but the Unruh
vacuum should be considered for an evaporating black hole.}

\subsection{Probing metric fluctuations}

Actually the contribution from small separation angles to the
integral discussed in the previous paragraph not only dominates the
integral but also gives a divergent result. This implies that the
event horizon is no longer well defined as a three-dimensional null
hypersurface when the effect of fluctuations is included since the
amplitude of its fluctuations is infinite. In order to get a finite
result, it is necessary to introduce some additional smearing along
the transverse null direction. The final result will then depend on
the characteristic size of the smearing functions employed. This
means that in addition to seeking workable prescriptions one should
also understand the physical meaning of the smearing introduced for
this task. This can be achieved by analyzing how the propagation of
test fields probes the metric fluctuations of the underlying
geometry.

For this purpose a natural first step is to study the effect of
those metric fluctuations on a bundle of null geodesics near the
horizon. This is expected to provide a good characterization of the
propagation of a test field whenever the geometric optics
approximation is valid \cite{BFP}, which is certainly the case when
studying Hawking radiation in the absence of fluctuations.  In
Ref.~\cite{BFP}, where the particular form of the metric
fluctuations was simply assumed rather than derived from first
principles, the authors found no dramatic effect due to the
fluctuations on the Hawking radiation associated with the test field
propagating in the black hole spacetime. In contrast, following the
stochastic gravity program,  an estimate of the metric fluctuations
exhibits a much more singular correlation function. Our preliminary
analysis suggests a larger effect on the propagation of a bundle of
null geodesics near the horizon, which would imply a substantial
modification of the Hawking effect derived under a test field
approximation.

However, it is likely that  the large fluctuations
found this way for a bundle of null geodesics sufficiently close to
the horizon is signaling a breakdown of the geometric
optics approximation rather than an actual drastic modification of
Hawking's result. This assessment stems from a similar situation in
flat space: Near a Rindler horizon large fluctuations arise from
using the geometric optics approximation, but they are artifacts
when examined in a more detailed quantum field theory calculation.
If confirmed, this would mean that the propagation of a bundle of
geodesics does not constitute an adequate probe of spacetime
fluctuations in this context. Although computationally more
involved, a more reliable way to extract physically meaningful
information about the effect of metric fluctuations on the Hawking
radiation is to consider the Wightman two-point function for the
test field, which characterizes the response of a particle detector
for that field, and analyze the radiative corrections when including
the interaction with the quantized metric perturbations.

%\newpage
\begin{acknowledgments}
We thank Paul Anderson, Larry Ford, Valeri Frolov, Ted Jacobson, Don
Marolf, Emil Mottola, Don Page, Renaud Parentani and Rafael Sorkin
for their interest in and useful discussions on various aspects of
this subject. This work is supported by NSF under Grant PHY-0601550.
A.~R.\ is also supported by LDRD funds from Los Alamos National
Laboratory.
\end{acknowledgments}

%\newpage

% \def\SBlongestlabel{A1} 
\SBtitlestyle{bar} 
\SBsubtitlestyle{none}

\end{document}